# How to hide your voice: Noise-cancelling bird photography blind

Caner Baydur[1†], Baojing Pu[2‡], Xiaoqing Xu[3§]

*1. Institute of Acoustics, School of Physics Science and Engineering, Tongji University, Shanghai 200092, China*

*2. Department of Landscape, School of Architecture and Urban Planning, Tongji University, Shanghai 200092, China*

**Abstract**

Getting close to birds is a great challenge in wildlife photography. Bird photography blinds may be the most effective and least intrusive way if properly designed. However, the acoustic design of the blinds has been overlooked so far. Herein, we present noise-cancelling blinds which allow photographing birds at close range. Firstly, we conduct a questionnaire in the eco-tourism centre located in Yunnan, China. Thus, we determine the birders' expectations of the indoor sound environment. We then identify diverse variables to examine the impact of architectural and acoustic decisions on noise propagation. Finally, we examine the acoustic performances of the blinds by considering the birds' hearing threshold. The numerical simulations are performed in the acoustics module of Comsol MultiPhysics. Our study demonstrated that photography blinds require a strong and thorough acoustic design for both human and bird well-being.

---

[†] canerbaydur@tongji.edu.cn, ORCID: 0000-0002-4207-1937.

[‡] 2130159@tongji.edu.cn, ORCID: 0000-0003-4810-5015.

[§] 17116@tongji.edu.cn, ORCID: 0000-0001-5908-0795, corresponding author.

# 1 Introduction

Ecotourism is defined as "responsible travel to natural areas that conserves the environment, sustains the well-being of the local people, and involves interpretation and education" according to the International Ecotourism Society (The International Ecotourism 2015). Under ideal circumstances, ecotourism develops the local economy and encourages the community to conserve natural resources (Sekercioglu 2002). However, keeping eco-tourism unobtrusive and sustainable is a challenge to achieve, as most of the environments in which birds inhabit, are areas with relatively intact but fragile ecosystems.

Ecotourists are often considered by birds to be predators (Frid and Dill 2002). Therefore, bird-based tourism brings along a range of adverse impacts on wildlife (Sekercioglu 2002; Slater et al. 2019). Birdwatching and photography are prominent activities where noise interferes with the habitat of birds. These outdoor activities are part of ecotourism and contribute to local economies (Sekercioglu 2002; Basnet et al. 2021). Bird photography varies from birdwatching. How? Bird photography is aimed at the satisfaction of visual evidence of birds (Manfredo et al. 1995). Three main reasons motivate bird photographers. Firstly, the conviction that the images will assist others to appreciate birds. Secondly, exploring rare or unusual birds. Lastly, the sense of accomplishment, or desire to educate, is inspired by sharing pictures with people (Slater et al. 2019). The negative effects of photography far outweigh those of birdwatching (Klein 1993; Tershy et al. 1997). In some circumstances, they may leave designated tracks and trails to take more satisfying photographs (Klein 1993). The equipment limitations also require bird photographers to reduce the distance to birds compared to birdwatchers (Lott 1992).

Birds are sensitive to noise according to sound frequencies and species (Beason 2004). Noise pollution that occurs in natural habitats may adversely affect their well-being substantially (Ortega 2012). Human-induced noise may significantly reduce bird populations (Putri et al. 2020; Bernat-Ponce et al. 2021), foraging opportunities (Navedo et al. 2019), nesting location (Zhang et al. 2017), and also alter the physiology of birds (Thiel et al. 2008). High noise levels can interfere with female birds' song-based assessment of male birds, leading to the female birds providing less energy and care to the chicks and eggs (Halfwerk et al. 2011). Auditory alertness of the species may be reduced if biological sounds are masked by anthropogenic sounds. In this case, the relationships between predator and prey are affected. Noise can cause delays in approaching and attacking prey (Halfwerk and van Oers 2020). Research indicates that a 5 dB increase in background sound levels means that prey can hear predators approaching at a 45% reduction in distance. Yet, it causes predators hunting with acoustic cues to experience a 70% reduction in the search area (Barber et al. 2009). Overall, the potential effects of noise on birds may be in a wide variety of aspects such as space use, communication, biology, reproduction, behaviour, ecosystem, and other (catch rate, genetic) (Sordello et al. 2020).

Photography blinds may be the most effective and least intrusive way to obtain high-quality and sharp images at close range. These structures conceal photographers, visually and audibly, from the habitat if properly designed. Nevertheless, most of the current bird photography facilities do not meet the acoustic requirements. Ma et al. (Ma et al. 2022) examined the effectiveness of bird blinds in mitigating birdwatcher impacts. The investigation is based on a field study in Hong Kong. As a consequence of the research, it has been shown that blinds can be an effective tool for visual disturbance, however, do not effectively reduce noise propagation. Researchers offer several comments aimed at noise reduction. For instance, creating a completely enclosed structure or using high soundproofing materials.

The use of sound insulation and absorbing materials in buildings is one of the major noise reduction methods. Sound insulation materials are used for cancelling noise transmission between two adjacent environments, which is assessed by its transmission loss, TL in dB (Arenas and Asdrubali 2018). For this purpose, numerous building materials and systems have been proposed so far (F. Alton Everest 1999). The mass of the materials can be increased to improve the insulation performance of the building elements (Schiavoni et al. 2016). If a multi-layer system is concerned, porous materials can be used, and/or an air gap can be created between layers. Thus, the insulation values of the system increase (Arenas and Asdrubali 2018). Sound-absorbing materials are also effective to decrease the noise level indoors (Cao et al. 2018), which is defined by its absorption coefficient, alpha, scaled from 0 to 1. As the absorption value of the material increases, the alpha approaches 1, and vice versa. The greater part of conventional sound-absorbing materials is based on porous and/or fibrous structures which provide high absorption of high-frequency sounds (Arenas and Asdrubali 2018). Microperforated panels ensure effective sound absorption at low- and middle-frequency regions (Maa 1998). The combination of microperforated panels and porous materials can be used to achieve broadband absorption in buildings (Duan et al. 2019; Shen et al. 2019). Although these materials provide an adequate noise reduction effect, to our knowledge, none of the research has applied these noise reduction methods to blinds so far. To apply these methods properly, comprehensive acoustic analyses of blinds should be conducted by considering various factors, including avian hearing threshold.

Here, we propose noise-cancelling bird photography blinds for improving the acoustic environment covering indoors and outdoors. To determine birdwatchers' expectations of the indoor sound environment, we conduct a questionnaire in the eco-tourism centre located in the Dahaizi region of Yunnan, China. We then propose blind design alternatives with architectural and acoustic variables aiming for noise reduction outdoors. The variables consist of building size, opened and closed windows, source loudness, and acoustic properties of building materials. The numerical simulations are performed in the acoustic module of Comsol MultiPhysics to estimate noise propagation. In the discussion, we examine the effects of the variables on acoustic propagation. The acoustic performance of the blinds is analysed by considering birds' hearing threshold. Finally, we discuss the acoustic challenges of blinds with the question, "How to prevent".

## 2 Materials and method

### 2.1 Research site and questionnaire

Until now, acoustic criteria for bird photography blinds have not been defined. For the subjective evaluation of the acoustic environment, there are many techniques, such as sampling strategies, the questionnaire, acoustic measurements, photos, video, environment information, and communication (Kogan et al. 2017). To determine the acoustic requirements of blinds, a questionnaire is conducted with 46 birders in the eco-tourism centre located in the Dahaizi region of Yunnan, China (27° 26′ 32″ N, 103° 19′ 18″ E) (Figure 1a). The questionnaire consists of two sections. The goal of the first part is to distinguish between annoying and pleasant sounds in the building. The second part is how much hear-abiotic, biotic, and anthropogenic sounds. In the light of the data, desired indoor sound environment in the blinds is determined.

The Dahaizi region is home to the endangered vulnerable black-necked cranes *Grus nigricollis* (De-Jun et al. 2011). Figure 1b illustrates the existing centre (red square), parking lot (black circle), highway (purple line), and birds' wintering area (area enclosed by blue). The photograph of the centre's immediate surroundings is available in Figure 1c, and its indoor environment in Figure 1d and Figure 1e.

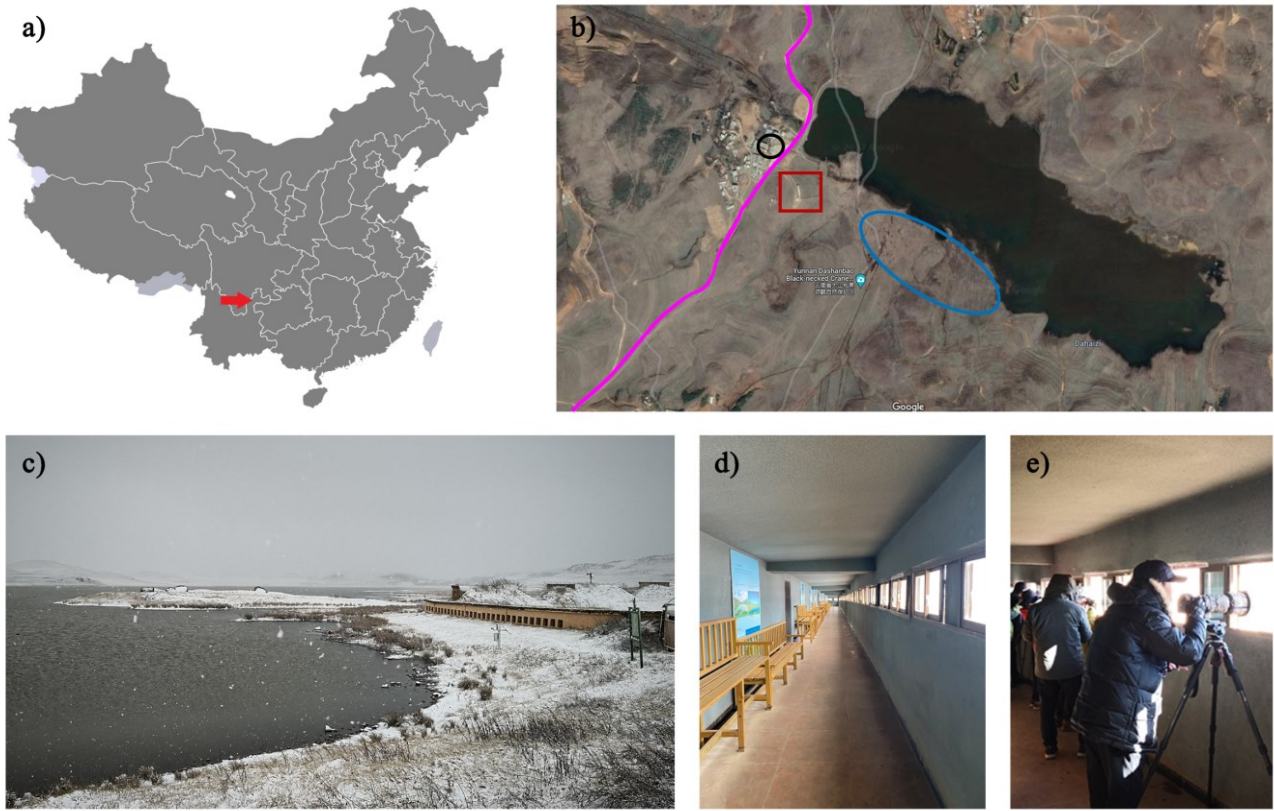

**Fig 1** (a) Location of the project site in China. (b) Location of the existing centre (Source: Google maps). (c) Immediate environment of the building. (d) Indoor view of the building. (e) People observing and photographing birds.

**2.2 Blind design with variables**

In this paper, we defined diverse variables to examine the effects on noise propagation. The variables include the materials' acoustic properties, source loudness, building size, open and closed window conditions.

**2.2.1 Material variables**

Here, we determine four scenarios to analyse the influence of materials' acoustic properties on the indoor and outdoor noise levels. The transmission losses of the materials in scenario 1 (SCN-1) are lower than in scenario 2 (SCN-2). The sound absorption coefficients of the materials in scenario 3 (SCN-3) are set to be lower than in scenario 4 (SCN-4). Thus, we can discuss to effects of the sound transmission losses on outdoor sound propagation by comparing SCN-1 and SCN-2. We can also examine the impacts of the sound absorption coefficients by comparing SCN-3 and SCN-4. The list of the scenarios is illustrated in Table 1. The sound absorption and insulation properties of the materials are given in the supplementary material Table S1 and Table S2, respectively.

**Table 1** Acoustic variables of the scenarios.

| Variable | Scenario Number | Wall - Ceiling | Door | Window |
|---|---|---|---|---|
| Sound insulation variables | SCN-1 | Hardboard | Hollow core | Ordinary glass |
| | SCN-2 | Single stud resilient channel wall | Solid timber door | Heavy glass |
| Sound absorption variables | SCN-3 | Unperforated wood | Unperforated wood | Ordinary glass |
| | SCN-4 | Perforated wood | Perforated wood | Ordinary glass |

**2.2.2 Loudness variables**

The human voice is considered as a noise source in the blind. In the simulation, the data from an experimental study is used for determining the sound characteristics of the talkers (Monson et al. 2012). The acoustic characteristic of voice may change depending on the speech volume, gender (Monson et al. 2012), and even language (Chu and Warnock 2002)! Diverse sound pressure levels (SPL) of the impact on the acoustics environment are investigated. The noise source is set at 54.8 dB for soft speech, 60 dB for normal speech, and 73.8 dB for loud speech (Monson et al. 2012).

**2.2.3 Building variables**

The building size and blind design with opened or closed windows are defined as the building variables. To show the effects of the building size on noise levels, two different building units which are small-sized and medium-sized are created. The small-sized (SS) blind (2.5m x 3m x 2.7m) has four windows and a 2-meter bench at 0.5-meter in width. The medium-sized (MS) blind (2.5m x 6m x 2.7m) has eight windows and a 4-meter bench. The size of each window is 0.5 in width and 0.4 in height. Whilst two noise sources are defined inside, as a maximum of two people can use the SS blind simultaneously, four people are considered for the MS blind. The sound absorption coefficient for the opened window is defined as 100% absorbent ($\alpha=1$) for all frequencies. The absorption coefficient of the closed window is given in the supplementary material Table S1, as "Ordinary glass" and "Heavy glass".

**2.2.4 Variable combinations**

Hereinafter, we systematically combine the aforementioned variables. Firstly, we investigate the impacts of the loudness on the acoustic environment with the analyses SS01, MS01, SS02, MS02, SS03, and MS03, by keeping window, transmission loss, and absorption coefficient conditions constant. Secondly, we examine the effects of the conditions of opened and closed windows with SS01/MS01 and SS04/MS04. Thirdly, we studied sound insulation and absorption variables with SS05/MS05 and SS06/MS06. We first combine materials with high sound-absorbing and low soundproofing, by using SCN-1 and SCN-4 simultaneously, with SS05/MS05. Vice versa, we then combine materials with low sound-absorbing and high soundproofing, by using SCN-2 and SCN-3 together, with SS06/MS06. Finally, we examine the acoustic environment with closed windows, high sound absorption, and insulation values with SS07/MS07. The list of variables for acoustic analyses is given in Table 2. "✓" refers to in use, whilst "✗" refers to out of use in the simulation.

**Table 2** List of variables with the numbers of the analyses.

| Analysis Number | SCN-1 | SCN-2 | SCN-3 | SCN-4 | Loud Speech | Normal Speech | Soft Speech | Opened window | Closed window |
|---|---|---|---|---|---|---|---|---|---|
| SS01/MS01 | ✓ | ✗ | ✗ | ✗ | ✓ | ✗ | ✗ | ✓ | ✗ |
| SS02/MS02 | ✓ | ✗ | ✗ | ✗ | ✗ | ✓ | ✗ | ✓ | ✗ |
| SS03/MS03 | ✓ | ✗ | ✗ | ✗ | ✗ | ✗ | ✓ | ✓ | ✗ |
| SS04/MS04 | ✓ | ✗ | ✓ | ✗ | ✓ | ✗ | ✗ | ✗ | ✓ |
| SS05/MS05 | ✓ | ✗ | ✗ | ✓ | ✓ | ✗ | ✗ | ✗ | ✓ |
| SS06/MS06 | ✗ | ✓ | ✓ | ✗ | ✓ | ✗ | ✗ | ✗ | ✓ |
| SS07/MS07 | ✗ | ✓ | ✗ | ✓ | ✓ | ✗ | ✗ | ✗ | ✓ |

## 2.3 Numerical model

Outdoor sound propagation may be predictable by using diverse methods including wave-based, ray tracing, and energy-based. So far, energy-based models were used for outdoor sound propagation, such as urban squares (Kang 2002, 2005), city street canyons (Okada et al. 2010), and complex urban environments with hybrid methods (Pasareanu et al. 2018). Numerical simulations are performed in the Acoustic Diffusion Equation Interface of the Comsol Multiphysics's Acoustics Module. The model is energy-based which means wave properties of sound are neglected. Since there are no obstacles such as walls and buildings in the outdoor environment, disregarding wave effects may be acceptable. Using diffusion-equation modelling (DEM) is a time-efficient way to get a result. Due to the advantages of DEM, all simulations were completed in just a few seconds. Furthermore, sound transmission losses of the materials can easily be defined in this interface of the software. Because our model was composed of diverse building elements and scenarios, this feature helped us to get results in a short time.

The SS and MS blinds are placed in a rectangular space (33.5m x 15m x 10m), as given in Figure 2a and Figure 2b. The noise sources are set at the position where a human observes birds (1.0 m from the ground, 0.5 m from the windows and 1.5 m from the walls), as shown in Figure 2b. A 30-meter receiver line, which starts from 1.5 meters away from the blind's façade is determined. The line is positioned at a height of 0.2 meters from the ground, considering the position of the birds on the ground during photography. Noise levels on this line are calculated. The line coloured red was illustrated in Figure 2a. The boundary condition of the rectangular space was set to simulate a free sound field as a perfectly sound-absorbing surface in the simulation.

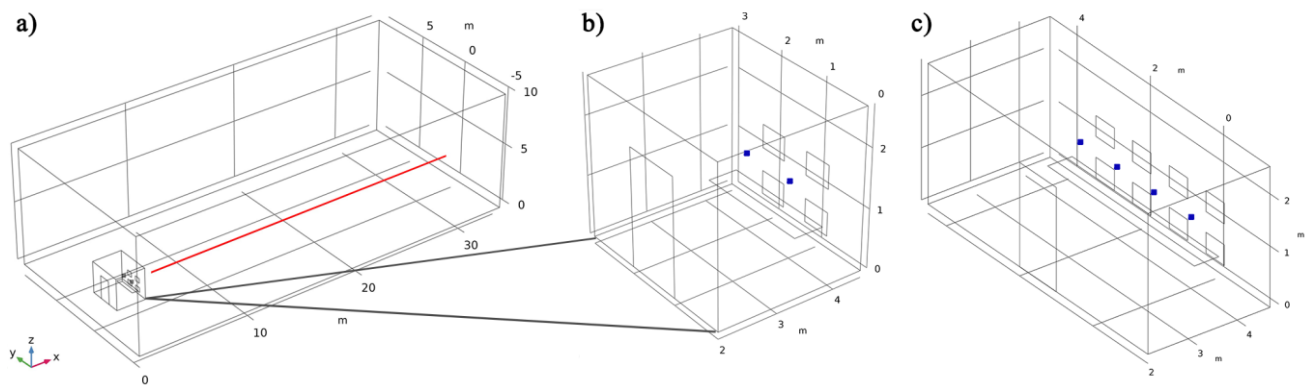

**Fig 2** (a) Comsol model of the blind with the outdoor environment. b) SS blind model and the positions of the sound sources (blue dot). c) MS blind model with the sources.

## 3 Results

### 3.1 Acoustic expectations of birders

The aim of the questionnaire is to determine people's expectations for the indoor acoustics environment when observing birds in a blind. Here in the questionnaire, we separated the audibility of the geological, biological, and anthropogenic sounds, as illustrated in Figure 3. In the section on geological sounds, wind sound was mostly heard, compared to others. Participants were undecided about the pleasure of this dominant sound. Some defined wind noise as pleasant, whilst others determined it as unpleasant. Other geological sounds, creek runnings and rain, were less audible but were noted as pleasant.

As biological sounds, almost all participants heard bird songs in the building. Similar to that ratio, people defined sounds as extremely pleasant. Although the poultry sounds were 32% audible, some participants defined them as pleasant whilst others as unpleasant. 11% of the birders heard wild animals (except birds) and described their sounds as pleasant.

On the part of anthropogenic noise, the audibility and pleasure range were parallel to each other. Approximately 55% of the birders heard sounds caused by several types of cars. Almost all participants defined those as unpleasant. Very few heard the guide's speech and those that did, described the speech as unpleasant. The sounds caused by tourists consisted of running, talking, shouting, and whispering. The most audible among them was "visitors talking" with 81%. It was followed by whispering at 59%. In the audibility graph, the difference between visitors talking and whispering is approximately 21%. On the other hand, in the pleasure graph, this difference increased to 34%. When visitors' conversations about dislike were evaluated, nearly half of these voices were defined as extremely unpleasant. The other half stated visitors' conversations as disturbing. However, looking at the whispering sounds, half of the participants described them as slightly disturbing. Only a few people rated it as extremely unpleasant.

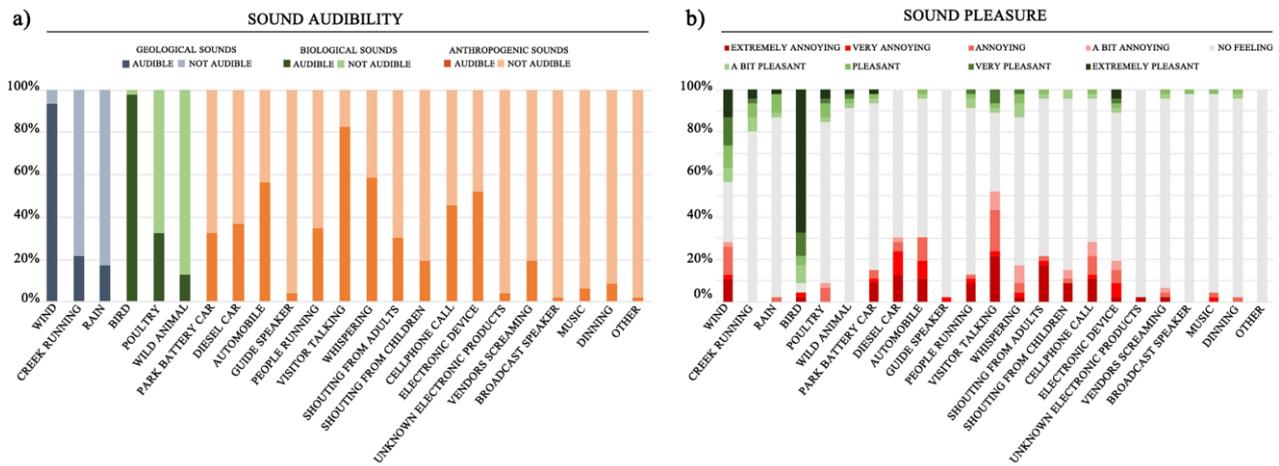

**Fig 3** (a) Sound audibility level. (b) Sound pleasure level.

As a result, the participants hear the 'visitor talking' as an anthropogenic sound at the highest ratio and are annoyed. Therefore, instead of a large-sized blind, we propose a small-sized design for a few people to meet the expectations of the birders.

### 3.2 Acoustic analyses

### 3.2.1 Building variables

Here, the effects of the building size with the source variables and the blind design with opened or closed windows on the noise level was investigated. Figure 4 (a) illustrates the SPL on the receiver line for the SS01, MS01, SS02, MS02, SS03, and MS03. The background noise level (BNL) was kept equal to a forest's sound level (Potočnik and Poje 2010a). The noise level transmitted from the MS blind is higher than that of the SS blind, about 3 dB more for all source variables. Figure 4 (b) indicates that the outside noise level is decreased by approximately 25 dB when the window is closed.

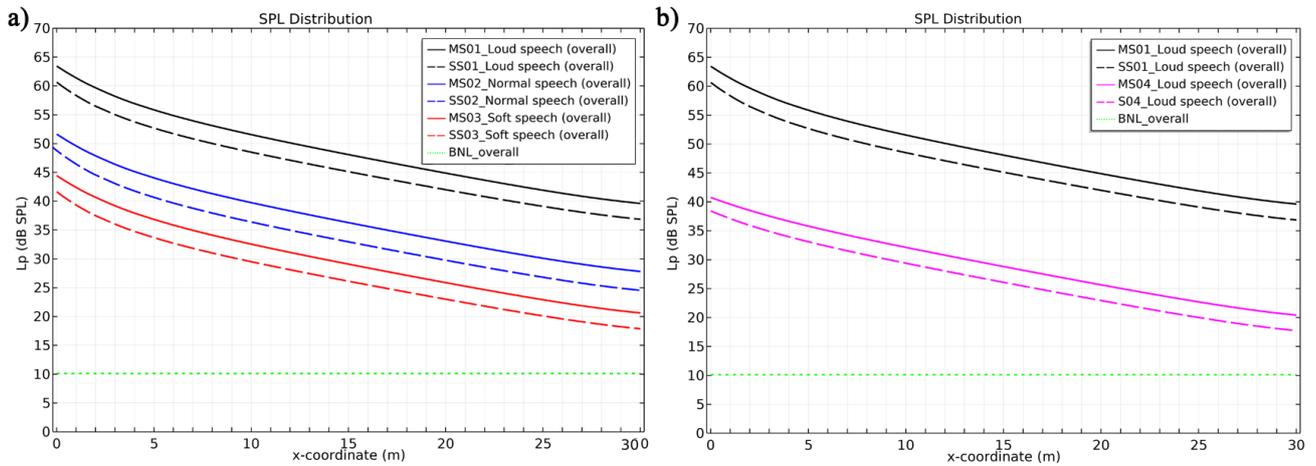

**Fig 4** SPL distribution (overall) of (a) the SS and MS blinds for loud, normal, and soft speech with the BNL for the forest environment (green dots), and (b) the SS and MS blinds with opened and closed windows for loud speech.

Figure 5 demonstrates the sound propagation path in the outdoor environment for the SS and MS blinds with the conditions of loud, normal, and soft speech. The noise level caused by the MS blind is higher than the SS blind on the x, y, and z-axes.

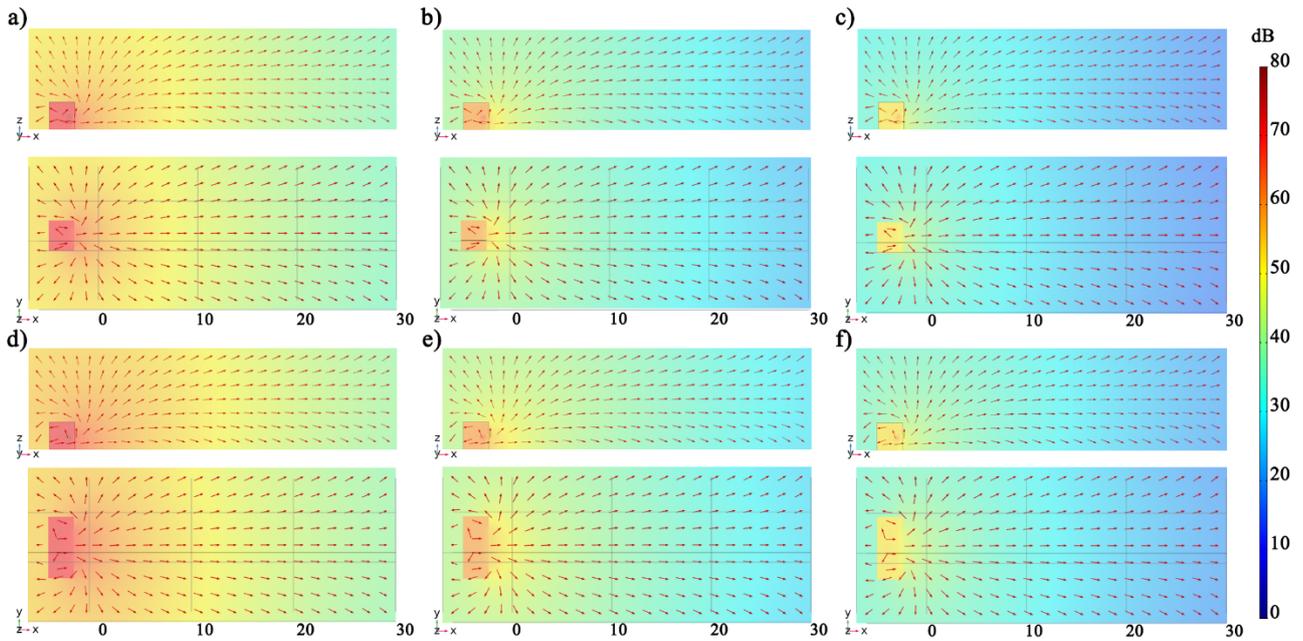

**Fig 5** Sound propagation path outdoor with the SS blind for (a) loud speech, (b) normal speech, and (c) soft speech. The propagation with the MS blind for (d) loud speech, (e) normal speech, and (f) soft speech.

### 3.2.2 Loudness variables

In this section, the impact of the source loudness on outdoor noise level is examined. The noise level emitted by loud-speaking bird watchers exceeds BNL at all frequencies for the SS and MS blinds, as illustrated in Figure 6 (a). In the 'normal speech', only sounds at 4000 Hz decrease to the same level as the background noise at 24 meters for the SS blind, as shown with a black asterisk in Figure 6 (b). In the 'soft speech', sounds of 4000 Hz at 12 meters, 2000 Hz at 11 meters, 125 Hz at 24 meters, and 1000 Hz at 27 meters for the SS blind (marked with a black asterisk) equal the relevant frequencies of BNL. Sounds of 2000 Hz and 4000 Hz at 16 meters for the MS blind (marked with a red asterisk) coincide with the relevant frequencies of BNL (Figure 6 (c)).

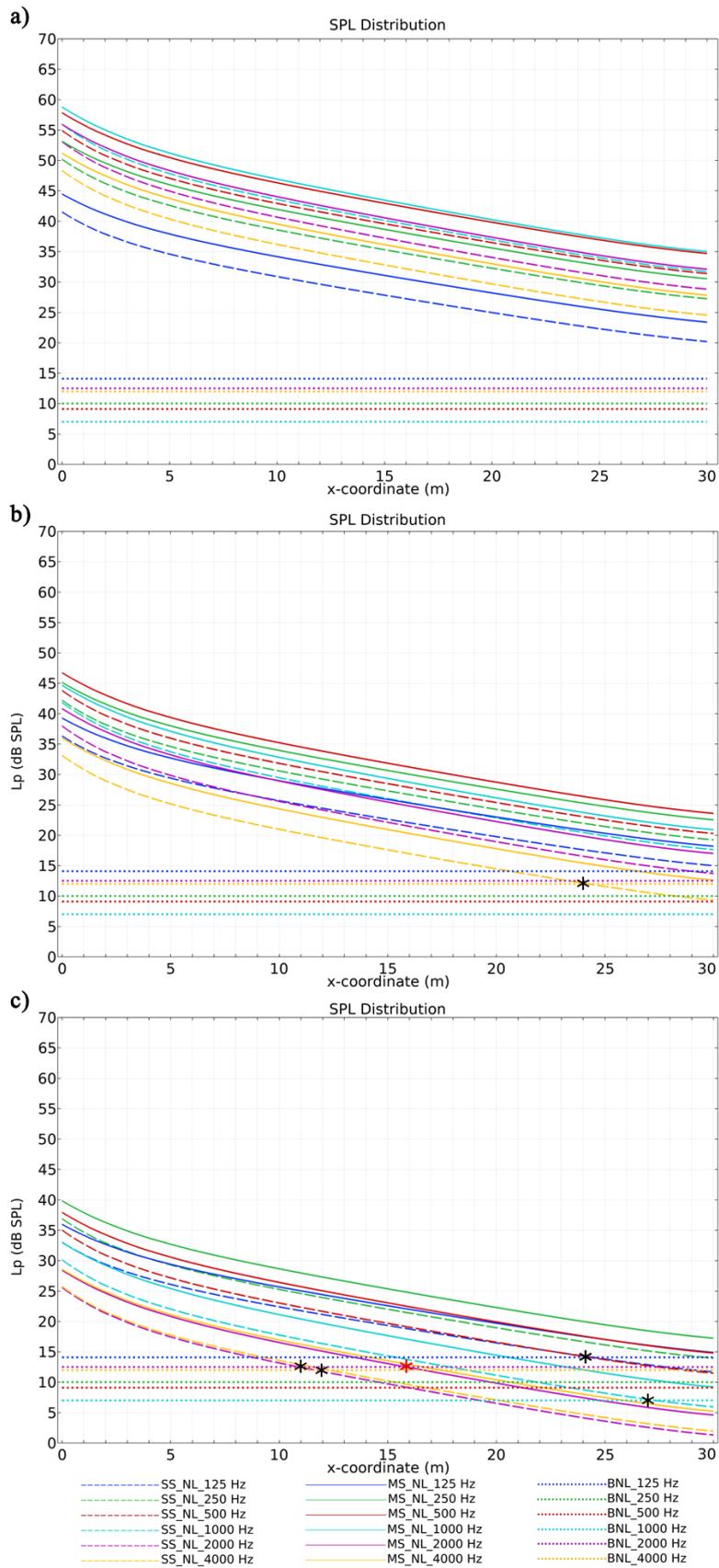

**Fig 6** Noise levels with the variables of (a) loud speech, (b) normal speech, and (c) soft speech depending on the frequency.

### 3.2.3 Material variables

We investigated the effects of the material variables on the outdoor noise level with SS04, MS04, SS05, MS05, SS06, MS06, SS07, and MS07. Figure 7 shows the comparison of the analyses with BNL. As we mentioned in Table 2, we demonstrated the acoustic analyses of closed window without an acoustic precaution with the SS04 and MS04. When the sound absorption coefficient of the materials in SS04/MS04, is improved, approximately 8 dB noise reduction is achieved with SS05 and MS05. If the sound transmission loss of the materials in SS04 and MS04 is increased, this reduction value increased to 15 dB with SS06 and MS06. In the case that the absorption and insulation coefficients of the materials were improved together, the noise was reduced by 23 dB with SS07 and MS07. The blinds achieved the same level as the background noise at the points which were marked asterisks. However, since these values are overall, it is crucial to examine the results on the frequency base which we will indicate in Figure 9.

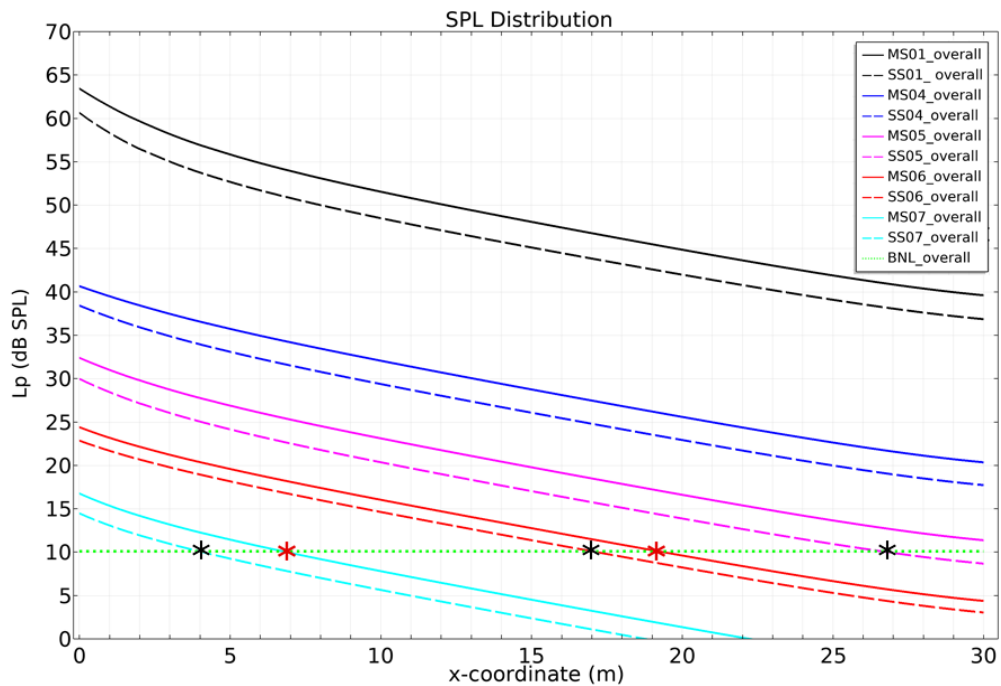

**Fig 7** Noise levels (overall) with the analyses number 01, 04, 05, 06, and 07 for SS and MS blinds including BNL (overall).

Figure 8 indicates the sound propagation path outdoor for the (a) SS04, (b) MS04, (c) SS05, (d) MS05, (e) SS06, (f) MS06, (g) SS07, and (h) MS04 with x-z and x-y axis.

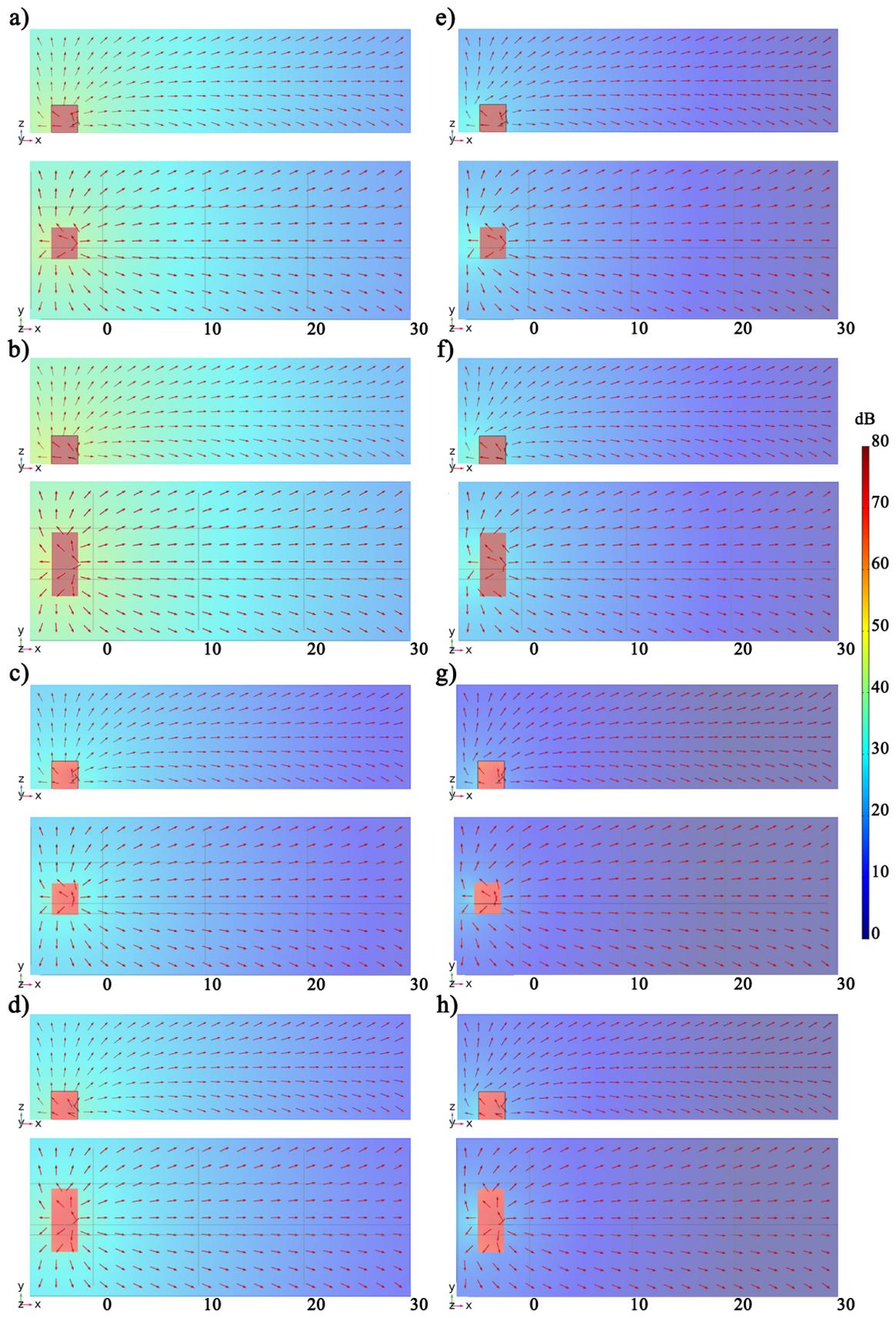

**Fig 8** Sound propagation path outdoor for the (a) SS04, (b) MS04, (c) SS05, (d) MS05, (e) SS06, (f) MS06, (g) SS07, and (h) MS04.

Figure 9 illustrates the noise levels with the variables of SS04, MS04, SS05, MS05, SS06, MS06, SS07, and MS07 with the BNL depending on the frequency. Figure 9 (a) indicates that the sound at 4000 Hz is equal to the same number of the BNL at 8 meters (SS blind) and 12 meters (MS blind) distance. For the SS blind, the sound of 2000 Hz at 23 meters is equal value to the relevant frequency of the BNL. All other frequencies exceed the background noise within the limits of the receiver line. When the absorption coefficients of the surface materials in the SS05 and MS05 blind were increased, the noise decreases at different levels at all frequencies (Figure 9 (b)). This reduction number may increase if the absorption coefficient of the materials improved on the basis of frequency. Table 2 shows the absorption coefficient of the wooden panel on the wall and ceiling was increased from 0.28 to 0.67 at 125 Hz, and from 0.11 to 0.96 at 4000 Hz. This leads that 36 dB noise at 125 Hz at 0 meters for SS blind was decreased to 34 dB. At the same point, the sound of 4000 Hz decreased from 20 dB to 14 dB. Therefore, a greater increase in the absorption coefficient of the material at 4000 Hz compared to the number at 125 Hz results in 4 dB more noise reduction is noticeable. For the MS blind, the noise level at 4000 Hz is equal to the same number of the BNL at 12 meters without absorbing material. The great improvement of the absorption coefficient led to a decrease in this distance to 2 meters. Similarly, the sound of 2000 Hz resulted in a significant distance reduction to the BNL level at the same frequency. Nevertheless, other frequencies exceeded the BNL.

Increasing the sound insulation value of the materials was in parallel with the frequency-based evaluation of the sound absorption coefficient. The SS06 and MS06 analyses in Figure 9 (c) showed that the materials with high transmission loss increased the noise reduction. All frequencies did not exceed the BNL after 18 meters for the SS blind, and after 21 meters for the MS blind. Therefore, after these distances, anthropogenic noise may not cause a problem. However, if photographing at close range (less than 18 meters) is desired, birds will be affected by noise as the sounds surpass the BNL.

For shooting close-up photography, we used materials with have high transmission loss and high absorption coefficient together. It greatly reduced the noise level that transfers from indoors to outdoors. Figure 9 (d) shows that after the 5 meters, all frequencies were below the BNL for the SS blind. For MS, this number was provided at 8 meters.

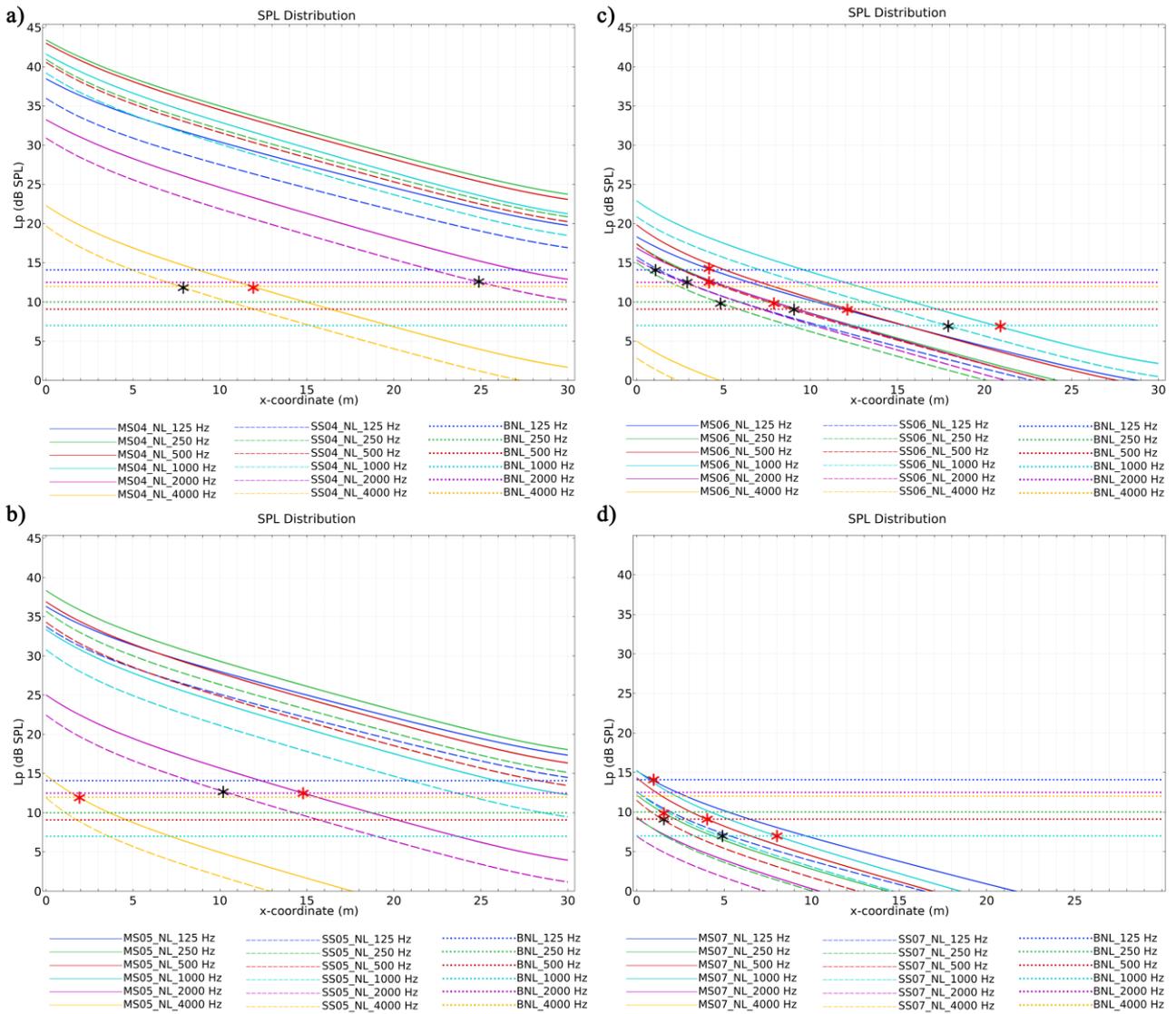

**Fig 9** Noise levels with the variables of (a) SS04, MS04, (b) SS05, MS05, (c) SS06, MS06, (d) SS07, and MS07 depending on the frequency.

**4 Discussion**

**4.1 Effects of the source loudness**

Our acoustic analyses of the loudness variable demonstrated that loud, normal, and soft speech leads to significant differences in noise levels. As shown in Figure 4 (a), there was a difference of 12 dB between high and normal speech, and 7 dB between normal and soft speech for the MS blind. This difference varied according to the frequency. Whilst there was about a 5 dB variance at 125 Hz between loud and normal speech, this number went up to 15 dB at 4000 Hz (Figure 6). Birds are more sensitive to high-frequency sounds than low frequencies (Beason 2004). Therefore, the increase in noise is remarkable. The noise source in our study is based on human speech. Nevertheless, other noise types may be considered in future studies. For instance, camera-shutter sounds may be examined as a source (Huang et al. 2011).

**4.2 Effects of the building size**

Aiming to reduce the anthropogenic noise both indoors and outdoors, we created the SS blind for two people and the MS blind for four people. When the building size was enlarged, the noise level increased due to the magnitude of people. The double noise source resulted in a 3 dB increase. Therefore, as can be seen in Figure 4 and Figure 7, the noise emitted from the MS blind was twice that of the SS blind. Consequently, large blinds require more high sound-absorbing and proofing materials compared to small blinds.

**4.3 Effects of the sound-absorbing materials**

Sound-absorbing materials may be a substantial solution since they reduce acoustic power in the enclosed (Adams 2016). With the increasing absorbing coefficient of the materials, the noise level was reduced by approximately 8 dB (overall), as shown in Figure 6. The noise reduction amounts varied as a result of the different increasing absorption coefficients of the materials on the basis of frequency (Figure 9 (a) and (b)). We used a sound-absorbing material which provides nearly perfect absorption (0.96) at 4000 Hz and a non-perfect absorption (0.67) at 125 Hz. The noise was reduced by about 7 dB at 4000 Hz, whilst a reduction of 2 dB was achieved at 125 Hz. The frequency range in which the sound-absorbing material is effective, is an important issue.

**4.4 Effects of the sound-proofing materials**

Sound-proofing materials are a major solution in cancelling noise from indoors to outdoors. Increasing the sound transmission loss of the materials (SS06) resulted in a noise reduction of about 16 dB (overall) compared with SS04, as shown in Figure 7. In the scenario where the absorption coefficient of the materials (SS05) is increased, the reduction amount was approximately 10 dB. When analysed on a frequency basis, the sound-proofing materials brought a reduction of 20 dB at 125 Hz and 17 dB at 4000 Hz (Figure 9 (a) and (c)). As a result, sound-proofing materials are more efficient than absorbing materials in our scenarios.

**4.5 Hearing characteristics of birds**

The evaluation of birds' hearing characteristics is vital in achieving a properly designed noise-cancelling blind. The noise level propagates from blinds must remain under the birds' hearing threshold. Avian hearing characteristics differ depending on their species. In general, small birds hear high-frequency sounds better, whilst large birds hear low-frequency sounds better (Köppl 2015). Among this diversity, some birds have remarkable hearing sensitivity. For instance, owls are known for their exquisitely sensitive hearing characteristics. They trust their capacity on hearing when hunting their prey at night (Köppl 2015). They are able to hear sounds below 0 dB SPL (Dooling et al. 2000). This means that their hearing threshold is lower than that of humans (Köppl 2015). Nevertheless, the average bird hearing threshold (ABHT) curve is higher and narrower than human's curve (Dooling and Popper 2007). As a rule of thumb, the frequencies of 1 to 5 kHz are the best hearing region for birds. The most sensitive regime is about 2 to 3 kHz with 0-10dB SPL (Dooling et al. 2000).

Figure 10 illustrates the noise-cancelling efficiency of the SS07 and MS07 blinds. The figure includes background noise level for a forest environment (Potočnik and Poje 2010a), ABHT (Dooling et al. 2000), and noise levels that propagate from the blinds at 0 meters.

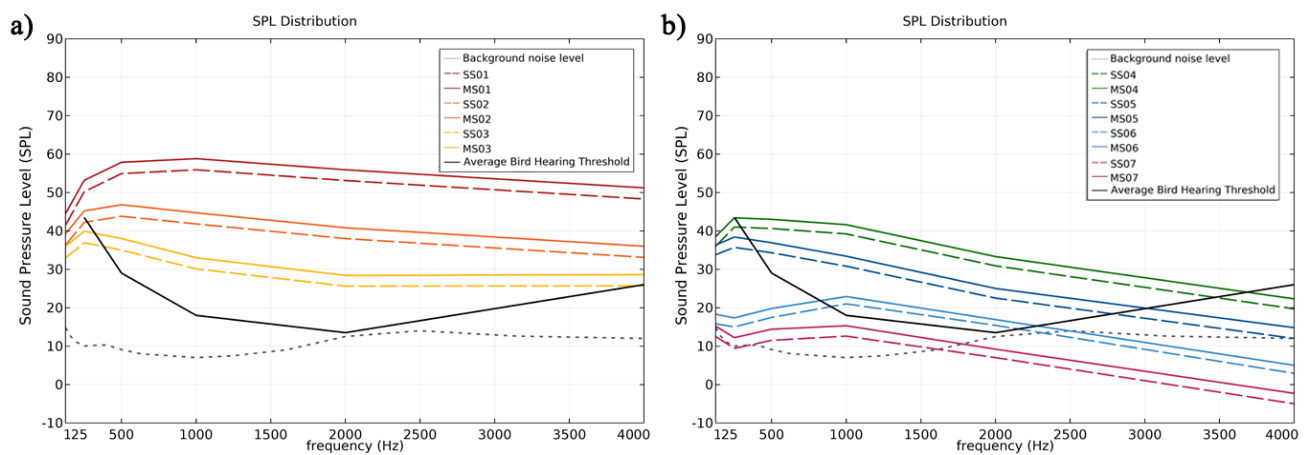

**Fig 10** Noise levels of (a) SS01, MS01, SS02, MS02, SS03, MS03 and (b) SS04, MS04, SS05, MS05, SS06, MS06, SS07, MS07 blinds with the forest environment (Potočnik and Poje 2010b) and ABHT (Dooling et al. 2000).

Figure 10 (a) illustrates that the transmitted noise level from the blinds with opened windows and without acoustic precautions, SS01, MS01, SS02, MS02, and SS03, MS03, remains above the ABHT. This means that birds hear photographers' noise between 125 Hz to 4000 Hz.

Figure 10 (b) indicates that when the windows are closed, but still do not add acoustic precautions, the noise propagated from SS04 and MS04 is higher than the ABHT. By keeping the windows closed, yet using sound-absorbing materials, the sound level at high frequencies (3000 Hz and above) is below the ABHT. When insulation materials are utilized instead of absorbers, the sound level is greatly reduced. However, the noise level is still high in the 1000-2000 Hz range which birds are sensitive to. The combination of insulation and absorbing materials, SS07/MS07, substantially enhances the noise reduction effect. Noise caused by SS07 and MS07 blinds is higher than the background noise level between 125 Hz to 1700 Hz. However, the ABHT curve is above the blinds' noise curve which means birds cannot hear birders' sounds. The blinds' noise remains under the background noise in frequencies between 1700 Hz to 4000 Hz. The ABHT is also higher than the noise levels of the background and blinds. Thus, noise transmitted from the blinds will not trigger and 'escape response'.

This comparison is based on the ABHT. Some birds' curve, such as owls, may be lower than the average bird's. According to the bird's hearing characteristics, the sound insulation and absorption properties of materials used in the blind, should adjust.

**4.6 How to reduce the noise spread caused by bird photographers efficiently?**

We recommend SS blinds which include a small number of people. The blind's window should be closed to decrease the outdoor noise level. Sound-absorbing materials reduced unwanted sounds indoors and outdoors. Sound-proofing materials were a 'strong tool' to eliminate noise outdoors. These materials can be used individually for situations where high noise cancellation is not needed, such as for long-distance photography. Nevertheless, for close-distance images, they have to be utilized simultaneously to minimize noise. We illustrated the exploded isometric structure for the SS07 in Figure 11.

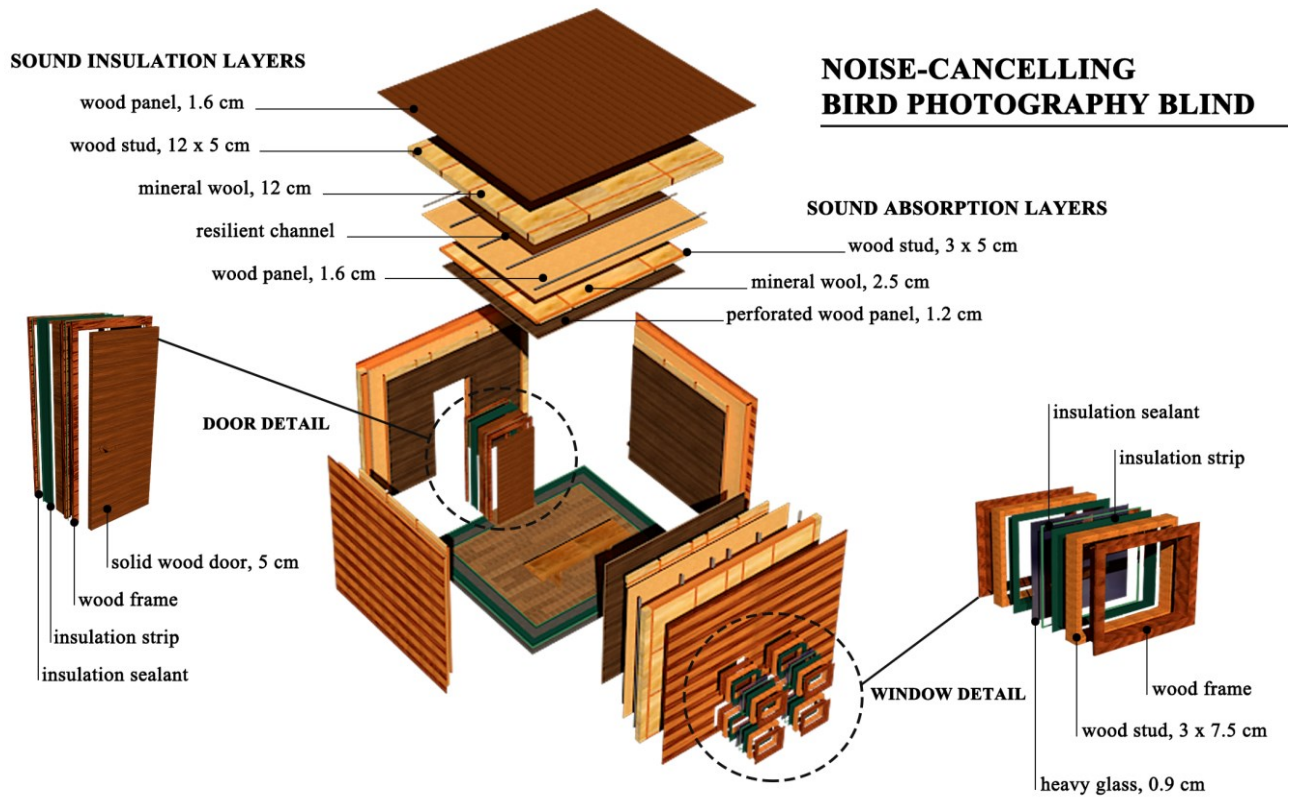

**Fig 11** Exploded 3D model of the SS07.

Additionally, evaluation of the data on the basis of the frequency is vital. Noise reduction in the overall frequency level may cause insufficient insulation. Birds generally hear high-frequency sounds better than low-frequencies, depending on the variety of species (Adams 2016). In a natural environment such as a forest, high-frequency sounds may be as low as 12 dB. The increase in speech loudness results in more rising in high-frequency sounds compared to low-frequencies. Choosing sound-absorbing and/or proofing materials which is efficient in the high-frequency range may be a proper solution.

## 5 Conclusions

Blinds conceal bird photographers, visually and audibly, from the habitat if properly designed. Thus, photographers at a close distance to birds can obtain high-quality and sharp images. Nevertheless, our results clearly demonstrated bird photography blinds require a strong and thorough acoustic design. In this paper, we presented noise-cancelling blinds which allow photographing birds at close range. We clearly demonstrated that the noise caused by the blinds remains under the birds' hearing threshold.

As a result of the analyses, we showed that small-sized blinds with closed windows may be a proper planning decision to reduce noise. Using sound-absorbing and proofing materials helped to decrease the outdoor noise level individually. We determined that using these materials simultaneously may be the best way to minimize noise both indoors and outdoors. Our analyses include various architectural and acoustic variables. However, further studies are necessary for diversifying these variables. Types of sound sources, background noise levels, thresholds of bird species, and various sound-absorbing and proofing materials can be expended. Thus, perfect noise-cancelling photography blinds can be achieved.

**Acknowledgements**

Caner Baydur would like to thank the China Scholarship Council (CSC) for their financial support during the duration of his master's courtesy the "Chinese Government Scholarship - Postgraduate Study Program" of Tongji University. We are grateful to Gulcin Konuk for providing helpful comments, and Jingjie Feng for helping visualisation of questionnaire results.

This study was part of a research project named "The construction of a multi-sensory perception system and its health effect in Jiuzhaigou World Heritage Site under the background of a post-epidemic situation under the background of a post-epidemic situation", and was supported by the Science and Technology Department of Sichuan Province (Grant No. 22GJHZ0142). Besides, the project was supported by the Science and Technology Commission of Shanghai Municipality with the project "Research on key technologies of intelligent management and control in the whole life cycle of large cultural parks" under the "Science and Technology Innovation Action Plan Social Development Science and Technology Project" Programme (Grant No. 21DZ1203004).

**Author Contribution**

Caner Baydur: Conceptualization, Investigation, Supervision, Methodology, Visualization, Software, Writing – original draft. Baojing Pu: Visualization, Writing - review & editing. Xiaoqing Xu: Conceptualization, Investigation, Supervision, Writing - review & editing, Funding acquisition.

**Declaration of competing interest**

The authors have no competing interests to declare that are relevant to the content of this article.

**Supplementary Material**

**Table S1** Sound absorption coefficients of the materials (Yang 2013; Cox and d'Antonio 2016; Everest and Pohlmann 2022).

| Element | Material | ASAC* | 125 Hz | 250 Hz | 500 Hz | 1000 Hz | 2000 Hz | 4000 Hz |
|---|---|---|---|---|---|---|---|---|
| Furniture | Wooden bench, a person | 0.76 | 0.57 | 0.61 | 0.75 | 0.86 | 0.91 | 0.86 |
| Floor | Linoleum on concrete | 0.03 | 0.02 | 0.03 | 0.03 | 0.03 | 0.03 | 0.02 |
| Wall and ceiling | Unperforated wood | 0.16 | 0.28 | 0.22 | 0.17 | 0.09 | 0.10 | 0.11 |
| Wall and ceiling | Perforated wood | 0.93 | 0.67 | 1.09 | 0.98 | 0.93 | 0.98 | 0.96 |
| Door | Hollow core door | 0.20 | 0.30 | 0.25 | 0.20 | 0.17 | 0.15 | 0.10 |
| Door | Solid timber door | 0.10 | 0.14 | 0.10 | 0.06 | 0.08 | 0.10 | 0.10 |
| Window | Ordinary glass | 0.17 | 0.35 | 0.25 | 0.18 | 0.12 | 0.07 | 0.04 |
| Window | Heavy glass | 0.06 | 0.18 | 0.06 | 0.04 | 0.03 | 0.02 | 0.02 |
| Outdoor wall and roof | Chipboard with mineral wool | 0.06 | 0.12 | 0.04 | 0.06 | 0.05 | 0.05 | 0.05 |
| Outdoor ground | Soil with vegetation | 0.76 | 0.39 | 0.68 | 0.78 | 0.94 | 0.95 | 0.83 |

* ASAC: Average Sound Absorption Coefficient

**Table S2** Sound insulation properties of the materials (Berendt and Cook 1958; Rudder 1985; Long 2005).

| Element | Material | ASTL* | 125 Hz | 250 Hz | 500 Hz | 1000 Hz | 2000 Hz | 4000 Hz |
|---|---|---|---|---|---|---|---|---|
| Wall and ceiling | Hardboard | 21 | 7 | 12.5 | 19 | 23 | 29 | 36 |
| Wall and ceiling | Single stud resilient channel wall | 46 | 30 | 43 | 49 | 49 | 52 | 56 |
| Door | Hollow core | 19 | 14 | 15 | 17 | 18 | 22 | 29 |
| Door | Solid timber door | 31 | 29 | 29 | 31 | 29 | 30 | 40 |
| Window | Ordinary glass | 25 | 20 | 21 | 25 | 26 | 31 | 30 |
| Window | Heavy glass | 35 | 29 | 34 | 35 | 35 | 34 | 45 |

* ASTL: Average Sound Transmission Loss